\documentclass[twocolumn]{cinc}
\usepackage[utf8]{inputenc}
\usepackage{graphicx}
\usepackage{amsmath} 
\usepackage{wrapfig}
\usepackage{multirow}
\usepackage{float}
\usepackage{subfig}
\usepackage{svg}
\usepackage{caption}
\usepackage{xurl}
\usepackage{algorithm}
\usepackage{program}
\usepackage{booktabs}
\usepackage{float}
\usepackage{multicol} 

\usepackage{enumitem}
\setlist[enumerate,1]{label= (\arabic*),leftmargin=*,align=right}

\begin{document}
\title{PhysioZoo: The Open Digital Physiological Biomarkers Resource}

\author{Joachim A. Behar$^{1}$, Jeremy Levy$^{1,2}$, Eran Zvuloni$^{1}$, Sheina Gendelman$^{1}$, Aviv Rosenberg$^{1,6}$, Shany Biton$^{1}$,  Raphael Derman$^{3}$, Jonathan A. Sobel$^{4}$, Alexandra Alexandrovich$^{1}$, Peter Charlton$^{5}$ and Márton Á. Goda$^{1}$\\
\ \\
$^1$Faculty of Biomedical Engineering, Technion, Technion–IIT, Haifa, Israel
\ \\
$^2$The Andrew and Erna Viterbi Faculty of Electrical \& Computer Engineering, Technion–IIT, Haifa, Israel
\ \\
$^3$Department of Anesthesiology, Rambam Medical Center, Haifa, Israel
\ \\
$^4$University of Geneva, Geneva, Switzerland
\ \\
$^5$University of Cambridge, Cambridge, United Kingdom
\ \\
$^6$ Faculty of Computer Science, Technion, Technion–IIT, Haifa, Israel
}
\maketitle

\begin{abstract}

PhysioZoo is a collaborative platform designed for the analysis of continuous physiological time series. The platform currently comprises four modules, each consisting of a library, a user interface, and a set of tutorials: (1) PhysioZoo HRV, dedicated to studying heart rate variability (HRV) in humans and other mammals; (2) PhysioZoo SPO2, which focuses on the analysis of digital oximetry biomarkers (OBM) using continuous oximetry (SpO2) measurements from humans; (3) PhysioZoo ECG, dedicated to the analysis of electrocardiogram (ECG) time series; (4) PhysioZoo PPG, designed to study photoplethysmography (PPG) time series. In this proceeding, we introduce the PhysioZoo platform as an open resource for digital physiological biomarkers engineering, facilitating streamlined analysis and data visualization of physiological time series while ensuring the reproducibility of published experiments. We welcome researchers to contribute new libraries for the analysis of various physiological time series, such as electroencephalography, blood pressure, and phonocardiography. You can access the resource at physiozoo.com. We encourage researchers to explore and utilize this platform to advance their studies in the field of continuous physiological time-series analysis.
\end{abstract}

\section{The PhysioZoo platform}
\begin{figure*}[h]
\centering
\includegraphics[width=0.95\textwidth]{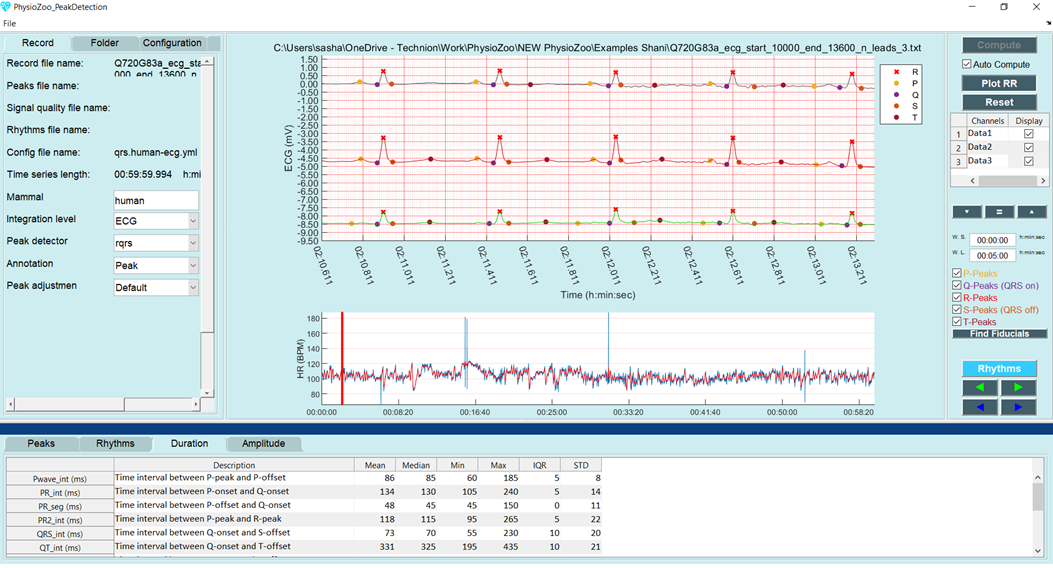}
\caption{PhysioZoo user interface. The interface enables to load and process a single recording. Fiducial points are detected and digital biomarkers are computed over the specified window. The specific example displayed is for a three lead ECG (Holter) recording using the pecg toolbox.}
\label{fig:pzmodulesinterface}
\end{figure*}

There is a notable lack of intelligent and robust algorithms that effectively utilize the valuable information embedded within physiological time series. This deficiency arises because of the limited availability of scientifically validated resources and the technical complexities associated with existing resources, mainly due to inadequate documentation and support. To address these challenges, we introduce PhysioZoo (PZ), an open digital physiological biomarker resource designed to facilitate research and unlock the potential wealth of information within physiological time series. Each PZ module offers essential functionalities, including prefiltering, handling of low-quality signals, precise fiducial point detection, and biomarker engineering. The primary objective of each module is to streamline the analysis of specific time series by providing a standardized pipeline, precise terminology, and a validated set of algorithms. Each module includes a code library and a user interface (Figure \ref{fig:pzmodulesinterface}).

PZ is a pioneering platform that recognizes the importance of tailoring the extraction of features to the unique characteristics of signals in different physiological modalities such as ECG, SpO2, and PPG. Unlike many existing toolboxes that claim to offer "universal" features applicable to any time series, which often fall short in practice, PZ embraces the body of domain-specific knowledge associated with each signal or modality. By incorporating mathematically defined biomarkers with physiological relevance, specific to the nature of the analyzed signal, PZ ensures the extraction of accurate and meaningful information. This approach allows researchers and clinicians, who generally have less technical skills but more domain knowledge, to gain insights into various physiological phenomena in a reliable and informative manner. 
Furthermore, PZ empowers researchers from nonbiological fields, who may lack the domain knowledge required to work with such data, to quickly extract meaningful features, e.g. for training machine learning models.


\begin{figure}
\centering
\includegraphics[width=0.5\textwidth]{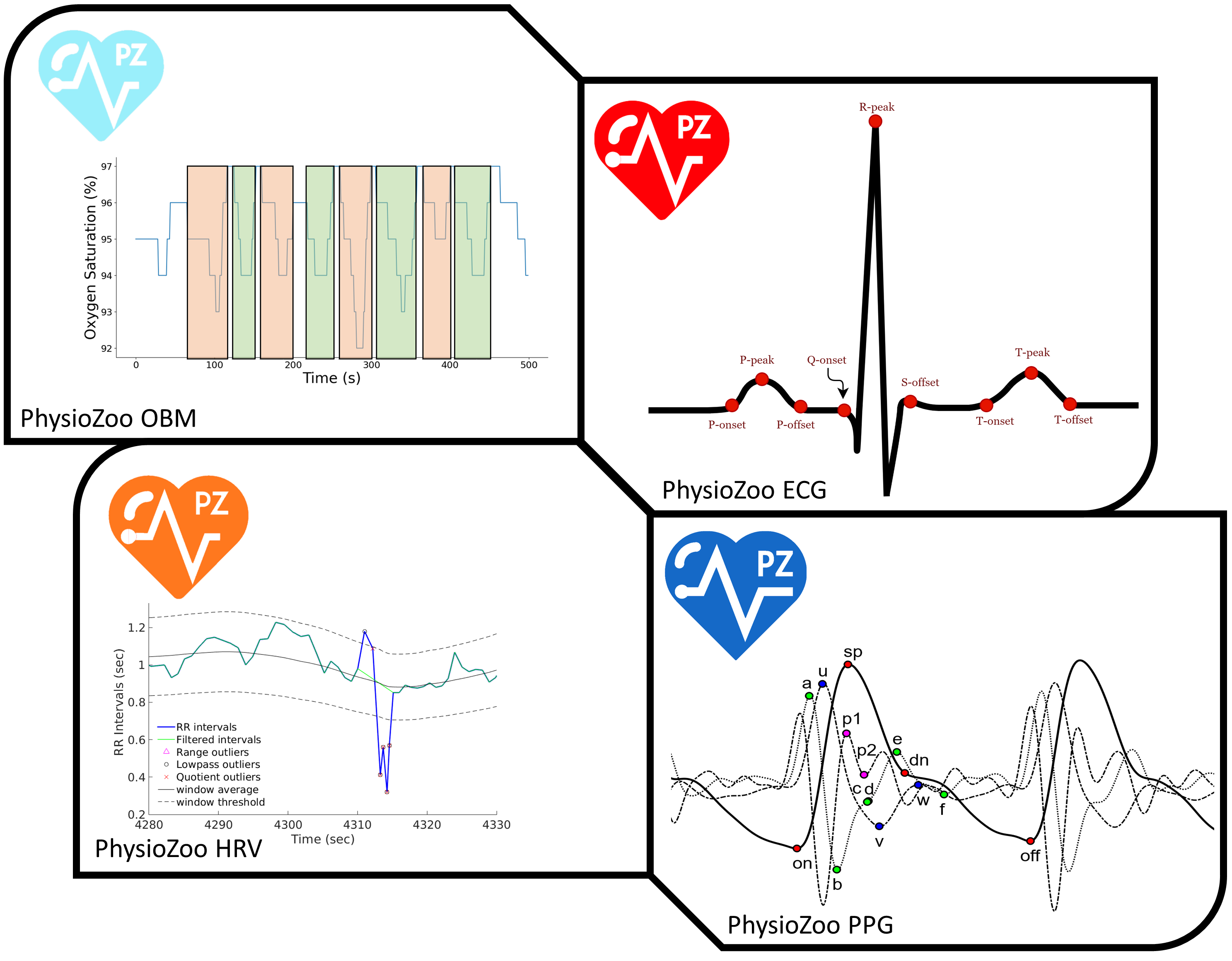}
\caption{The four modules available in PhysioZoo (PZ) enabling the processing of beat-to-beat, electrocardiogram (ECG), oximetry and photoplethysmography (PPG) time series. HRV: heart rate variability; OBM: digital oximetry biomarkers.}
\label{fig:pzmodules}
\end{figure}

Being an open-source initiative, PZ is committed to transparency and accessibility. Our user-centered approach incorporates intuitive toolboxes and a user interface that simplifies data visualization and ensures accessibility for non-technical users to analyze data samples effectively. The PZ modules are compatible with various data formats, such as CSV, txt, and the PhysioNet WFDB format, allowing seamless integration with diverse datasets. Furthermore, PZ operates under the GNU General Public License, as published by the Free Software Foundation, fostering a collaborative and open community dedicated to advancing the field of physiological time-series analysis. 

Each module is independently structured, documented, and managed by its original authors. This decentralized approach offers several advantages, empowering new contributors to take ownership of their projects while ensuring that all libraries remain part of an overarching initiative following a set of principles and scientific standards concerning the quality control of implemented functions. This includes validating source code against available benchmarks and / or annotated datasets, allowing the code to be open source, providing comprehensive documentation and the availability of a user interface. A set of video tutorials is available on the PZ YouTube channel (physiozoosoftware62).

\section{PhysioZoo modules}

A total of four PZ modules have been released (Figure \ref{fig:pzmodules} and Table \ref{table:1}) \cite{behar2018physiozoo,levy2021digital,gendelman2021physiozoo}. The inaugural PZ module, PZ HRV, was launched in 2018, marking the beginning of this initiative. PZ HRV, focuses on studying heart rate variability (HRV) in both humans and other mammals. The second module, PZ SPO2, is dedicated to the analysis of digital oximetry biomarkers (OBM) derived from continuous oximetry (SpO2) measurements in humans. The third module, PZ ECG, is designed for the analysis of electrocardiogram (ECG) time series data. Lastly, the fourth module, PZ PPG, is centered on studying photoplethysmography (PPG) time series. The release of these modules represents a significant step forward in the pursuit of advancing physiological time series analysis.


\textbf{PZ HRV \cite{behar2018physiozoo}.} Heart rate variability, the variation of the time interval between heartbeats contains important information about cardiac health and the autonomic nervous system \cite{rajendra2006heart,faust2022heart}. HRV can be used to quantify the short- and long-term dynamics of heartbeats. A total of 26 HRV biomarkers were implemented as part of PZ HRV, in the mhrv toolbox. The biomarkers are divided into three standard categories from the HRV literature: time domain, frequency domain, and non-linear. We contributed a standardized, validated HRV toolbox, which can be used not only to analyze human interval data but also comes with configuration presets supporting the analysis of data from other mammals (dogs, rabbits, mice). The mhrv toolbox is implemented in MATLAB.

\textbf{PZ SPO2 \cite{levy2021digital}.} Pulse oximetry is routinely used to non-invasively monitor oxygen saturation levels. A low oxygen level in the blood means low oxygen in the tissues, which can ultimately lead to organ failure. The biomarkers were divided into five categories: (1) General statistics: time-based statistics describing the SpO2 data distribution, (2) Complexity: quantifies the presence of long-range correlations in non-stationary signal, (3) Periodicity: quantifies consecutive events to identify periodicity in the SpO2 signal, (4) Desaturations: time-based descriptive measures of the desaturation patterns occurring throughout the signal, and (5) Hypoxic burden: time-based measures quantifying the overall degree of hypoxemia imposed on the heart and other organs during the recording period. A total of 44 digital oximetry biomarkers (OBM) were implemented in the pbom toolbox. Studying the variability of the continuous oxygen saturation time series using pbom may provide information on the underlying physiological control systems and enhance our understanding of the manifestations and etiology of diseases, with an emphasis on respiratory diseases. The toolbox, pobm, is implemented in Python.

\textbf{PZ ECG \cite{gendelman2021physiozoo}.} The electrocardiogram, or ECG, is a widely-used tool used in medical practice to identify cardiac disease. We implemented clinically important digital ECG biomarkers in the pecg toolbox to create a reference toolbox for ECG morphological analysis. The library and associated user interface can be used to process 12-lead ECG as well as long-term recordings, for example, from Holter monitors. A total 22 biomarkers were implemented. These were divided into two categories: (1) interval and segments, and (2) wave characteristics. The library, pecg, is implemented in Python, but uses the MATLAB wavedet \cite{martinez2004wavelet} package to extract a set of biomarkers based on the location of the fiducial points. This choice, although non-ideal, was made because we preferred to leverage a validated algorithm for fiducial points detection versus open source Python algorithms that were not validated. 

\textbf{PZ PPG \cite{goda2023physiozoo}.} PPG is an optical sensing technique widely used to monitor health and fitness in clinical and consumer devices \cite{Charlton2020rev}, such as smartwatches and pulse oximeters. PPG signals contain a wealth of information about the function of the heart, blood vessels, breathing, and the autonomic nervous system \cite{Allen_2007}. The PPG biomarkers are classified into four groups: (1) PPG Signal: biomarkers based on the location of the fiducial points of the PPG signal; (2) Signal Ratios: biomarkers based on ratios of the fiducial points of the PPG signal; (3) PPG Derivatives: biomarkers based on the location of the fiducial points of the PPG derivatives; and (4) Derivatives Ratios: biomarkers based on ratios of the fiducial points of the PPG derivatives. A total of 74 PPG digital biomarkers were implemented in the pyPPG toolbox. The pyPPG toolbox also provides general fiducial point detection for PPG signals.

\begin{table}
\begin{center}
\begin{tabular}{ |c|c|c|c| } 
 \hline
 Module  & Library & Language & Release \\ 
 \hline
 PZ HRV  & mhrv   & MATLAB & 2018 \\ 
 \hline
 PZ SPO2 & pobm   & Python & 2021\\ 
 \hline
 PZ ECG  & pecg  & MATLAB, Python & 2023\\ 
 \hline
 PZ PPG  & pyPPG  & Python & 2023\\ 
 \hline
\end{tabular}
\end{center}
\caption{PhysioZoo modules, libraries, implementation language and year of the first public release. The libraries and user interface are available at physiozoo.com}
\label{table:1}
\end{table}

\section{Discussion and future work}
Within the field of biomedical signal processing, much work has been done, through initiatives such as PhysioNet \cite{goldberger2000physiobank},  to make datasets openly available, standardized, and interoperable. Comparatively, less effort has been invested in making analytical tools openly available. Although there are some existing tools for each signal modality, they may often lack scientific validation or comprehensive documentation and generally do not have the benefit of a cohesive framework, which PZ provides. Thanks to the researchers who have contributed to PZ to date, four modules have been released. These enable the streamlined and reproducible analysis of HRV, ECG, SpO2, and PPG time series. PZ modules have been used in a number of scientific studies \cite{aublin2022predict, zvuloni2023merging, zvuloni2022atrial, biton2023generalizable, alkhodari2020predicting, albuquerque2020wauc, rosenberg2020signatures, levy2023deep, sobel2023descriptive, ben2023positional}.

There are two main limitations of our efforts. The first is based on the mixture of MATLAB and Python across the constituents of the PZ platform. This is currently due to depending on the original wavedet algorithm, the mhrv toolbox, and the PZ graphical user interface, all of which are implemented in MATLAB. An effort to convert each module's codebase to Python is needed. The second limitation is the need for engagement of new independent contributors to the platform. We encourage scientists and engineers to recognize the great potential of open source scientific tools to contribute to the growing set of PZ modules. In addition, there is ample opportunity to contribute new functions to the existing modules and toolboxes.  Some ideas include detection of arrhythmias, cardiac abnormalities, abnormal beats in ECG or PPG, respiratory rate estimation and respiratory signal analysis or algorithms for electroencephalogram time series analysis. Finally, by contributing their existing software as a new PZ module or into an existing module, researchers can also benefit from wider exposure and user-base for their tools. We thus encourage researchers, scientists and engineers to utilize and contribute to this resource, thereby facilitating open-source collaboration and high standards of research within the domain of physiological time series analysis.

\bibliographystyle{cinc}
\bibliography{new_refs}

\vspace{-0.2cm}
\begin{correspondence}
PhysioZoo admin (physiozoolab@gmail.com)\\
\end{correspondence}

\end{document}